\documentclass{elsart}
\usepackage{natbib}
\usepackage{graphicx}
\begin{document}
\runauthor{Y.~Aritomo and M.~Ohta}
\begin{frontmatter}
\title{Origin of the drastic decrease of fusion probability in
superheavy mass region}
\author[Dubna]{Yoshihiro Aritomo}
\author[Kobe]{Masahisa Ohta}

\address[Dubna] {Flerov Laboratory of Nuclear
Reactions, JINR, Dubna, Russia}
\address[Kobe]{Department of Physics, Konan University, 8-9-1
  Okamoto, Kobe, Japan}

\begin{abstract}
The fusion-fission process in the superheavy mass region is
studied systematically by solving the time evolution of nuclear
shape in three-dimensional deformation space using the Langevin
equation. By analyzing the trajectory in the deformation space, we
identify the critical area when the trajectory's destination is
determined to be the fusion or the quasi-fission process. It is also
clarified that the potential landscape around the critical area is
crucial for estimating the fusion probability, and its dependence
on the atomic number is presented.
\end{abstract}
\end{frontmatter}


\section{Introduction}

The fusion reaction in which two nuclei combine shows a variety of
phenomena controlled by properties inherent in light to heavy
nuclei. In the fusion reaction in which a compound nucleus with
atomic number $Z < 70$ is formed, the cross section of the
reaction can be described by the critical distance model
\cite{glas74} based on the strong absorptive nature of nuclei when
they approach each other to within the contact distance where the
configuration is more compact than that of the saddle shape for
fission. For the region of atomic number $70 < Z < 90$, however,
the critical distance model breaks down due to the effect of
dissipation force near the contact area, and therefore the
extra-push model is proposed \cite{swia81,bjor82}. The term {\it
fusion hindrance} is focused on this extra-push model. In addition
to dissipation, when the atomic number increases beyond 100, the
geometrical inversion between the contact point of two nuclei and
the saddle point of the composite system substantially affects the
fusion probability \cite{bloc86,nix77}. Therefore, since the
saddle point is located inside the contact point, it is expected
that the essential factor for fusion hindrance is strongly related
with the landscape of the region in which two nuclei are
considerably overlapping.

In this paper, we show how the fusion hindrance should be
described in heavy and superheavy nuclei with the region $Z >
100$. In the region, due to the strong Coulomb repulsion force,
the shape of the fragment of fissioning nucleus is easily deformed
in the fusion-fission process. Therefore, it is extremely
important to take into account the deformation of the fragments.
The fusion probability in the superheavy mass region has already been reported in references \cite{ohta03,ari04,ari05} taking account of the fragment deformation. On the basis of the results of our previous studies, we here present the origin of the fusion hindrance as determined by the analysis of the dynamical evolution
of nuclear shape during the fusion process.


Generally, the stability for fission in heavy nucleus is discussed
on the fission barrier height, which is calculated, for example, by the liquid
drop model and the shell correction energy. In the liquid drop part,
we can understand the general tendency of the fission life time
which decreases exponentially with increasing the atomic number of
the nucleus. On the other hand, we can understand the irregularity
of the nuclear property by the shell correction energy, and
explain the enhancement of the stability of the superheavy mass
nucleus. In the same way, the fusion process should be discussed both in the
macroscopic point of view and the microscopic one.
Even though the fusion barrier is modified by the shell effect, the extent of the fusion enhancement due to the shell effect is depends on the system, namely the potential landscape around the contact point. This is still an open problem and in this paper we restricted to discuss the major part of the fusion hindrance coming from the macroscopic potential. However, we remark that how the shell effect modify the trend of the hindrance due to the macroscopic potential.



In section~2, we briefly explain our framework for the study and
the model. We investigate the fusion hindrance precisely in
section~3. In this section, we discuss the stability of the
deformation in superheavy mass nuclei, and present how the fusion
hindrance appears with increasing the atomic number of the
colliding partner, using the mean trajectory calculation. The
critical condition of the fusion process is discussed. Fusion
probability is calculated by the three-dimensional Langevin
equation and the role of the shell effect in the fusion process is
discussed. In section 4, we present a summary and further
discussion to clarify the reaction mechanism in the superheavy
mass region.


\section{Model}

Using the same procedure as described in reference \cite{ari04},
to investigate the fusion-fission process dynamically, we use the
fluctuation-dissipation model and employ the Langevin equation for
the estimation of fusion probability. We adopt the
three-dimensional nuclear deformation space given by two-center
parameterization \cite{maru72,sato78} and the time evolution of
the nuclear shape is calculated by Langevin equation (we call it
{\it trajectory} in the deformation space). The three collective
parameters involved in the calculation are as follows: $z_{0}$
(distance between two potential centers), $\delta$ (deformation of
fragments) and $\alpha$ (mass asymmetry of the colliding nuclei);
$\alpha=(A_{1}-A_{2})/(A_{1}+A_{2})$, where $A_{1}$ and $A_{2}$
denote the mass numbers of the target and the projectile,
respectively. We assume that each fragment has the same
deformation as the first approximation. The neck parameter
$\epsilon$ is defined in the same manner as reference
\cite{maru72}. In the present calculation, $\epsilon$ is fixed to
be 1.0, so as to retain the contact-like configuration more
realistically for two-nucleus collision. The multidimensional
Langevin equation is given as

\begin{eqnarray}
\frac{dq_{i}}{dt}&=&\left(m^{-1}\right)_{ij}p_{j},\nonumber\\
\frac{dp_{i}}{dt}&=&-\frac{\partial V}{dq_{i}}
                 -\frac{1}{2}\frac{\partial}{\partial q_{i}}
                   \left(m^{-1}\right)_{jk}p_{j}p_{k}
                  -\gamma_{ij}\left(m^{-1}\right)_{jk}p_{k}
                  +g_{ij}R_{j}(t),
\end{eqnarray}
where a summation over repeated indices is assumed. $q_{i}$
denotes the deformation coordinate specified by $z_{0}$, $\delta$
and $\alpha$. $p_{i}$ is the conjugate momentum of $q_{i}$. $V$ is
the potential energy, and $m_{ij}$ and $\gamma_{ij}$ are the
shape-dependent collective inertia parameter and dissipation
tensor, respectively. A hydrodynamical inertia tensor is adopted
in the Werner-Wheeler approximation for the velocity field, and
the wall-and-window one-body dissipation is adopted for the
dissipation tensor \cite{bloc78,nix84,feld87}. The normalized
random force $R_{i}(t)$ is assumed to be  white noise, {\it i.e.},
$\langle R_{i}(t) \rangle$=0 and $\langle R_{i}(t_{1})R_{j}(t_{2})
\rangle = 2 \delta_{ij}\delta(t_{1}-t_{2})$. The strength of
random force $g_{ij}$ is given by $\gamma_{ij}T=\sum_{k}
g_{ij}g_{jk}$, where $T$ is the temperature of the compound
nucleus calculated from the intrinsic energy of the composite
system.
The potential energy is defined as

\begin{equation}
V(q,l)=V_{DM}(q)+\frac{\hbar^{2}l(l+1)}{2I(q)},
\end{equation}
\begin{equation}
V_{DM}(q)=E_{S}(q)+E_{C}(q),
\end{equation}
where $I(q)$ is the moment of inertia of a rigid body at
deformation $q$
$V_{DM}$ is the potential energy of the finite-range liquid drop
model. $E_{S}$ and $E_{C}$ denote a generalized surface energy
\cite{krap79} and Coulomb energy, respectively. The centrifugal
energy arising from the angular momentum $l$ of the rigid body is
also considered. The detail is explained in reference
\cite{ari04}.






\section{Origin of the fusion hindrance in superheavy mass region}

\subsection{Role of the nuclear deformation}


Figure~1 shows the potential energy surface of the liquid drop
model ($LDM$) for $^{224}$Th~(left) and $^{292}$114~(right) on the
$z-\delta$ plane with angular momentum $l=0$ and symmetry
$\alpha=0$. This potential energy surface is calculated using the
two-center shell model code \cite{suek74,iwam76}. The contour
lines of the potential energy surface are drawn in steps of 2~MeV.
As described in reference \cite{ari04}, to save computational
time, we use scaling and employ the coordinate $z$. The coordinate
$z$ is defined as $z=z_{0}/(R_{CN}B)$, where $R_{CN}$ denotes the
radius of the spherical compound nucleus. Parameter $B$ is defined
in terms of the fragment deformation parameter $\delta$ as
$B=(3+\delta)/(3-2\delta)$. In Fig.~1, the position at
$z=\delta=0$ corresponds to a spherical compound nucleus marked by
the circle $(\circ)$.


\begin{figure}
\centerline{
\includegraphics[height=.38\textheight]{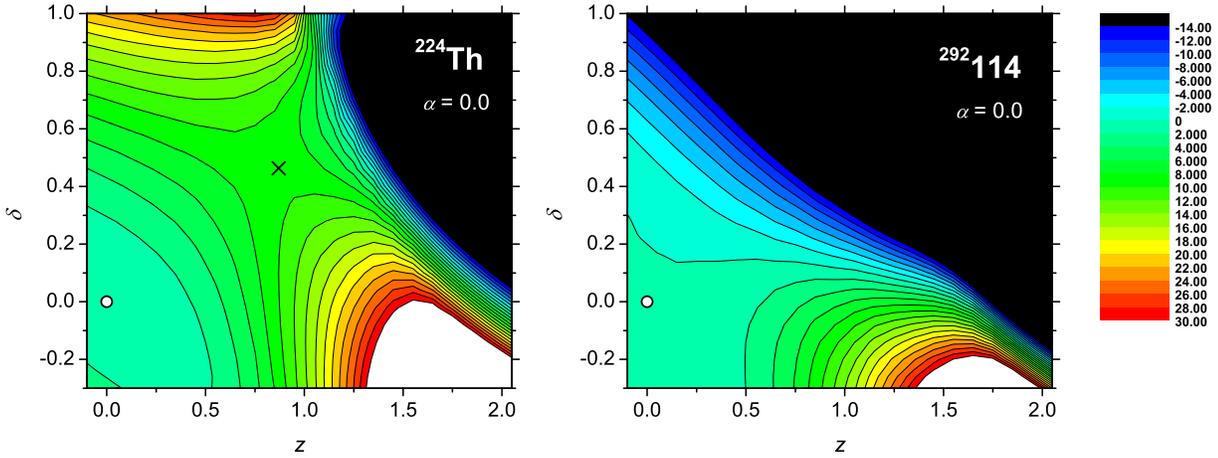}
}
  \caption{The potential energy surface
of a liquid drop model on the $z-\delta$ plane for $^{224}$Th
(left) and $^{292}$114 (right). The mass asymmetric parameter
$\alpha$ is zero. The saddle point marked by $\times$ is well
observed in the case of $^{224}$Th . The stability of the
spherical nucleus, marked by circles, against the parameter
$\delta$ can be seen for Th but not for the nucleus with $Z$=114.}
\end{figure}


For $^{224}$Th, we can see the pocket located in the spherical
region and the nucleus in this pocket is rather stable being
protected by the well-defined fission barrier marked by the cross
$(\times)$. On the other hand, for $^{292}$114, no fission barrier
can be seen when the shell effect is not taken into account, and
the nucleus around the spherical region is unstable against
fragment deformation, as shown in Fig.1, due to the Coulomb force
acting between two centers. That is, the system tends to rupture
easily, which induces fragment deformation competing with the
shell effect. The instability of the superheavy nucleus has been
discussed from the viewpoint of the fission barrier due to the
shell effect along the $z$-direction, but we have never paid
attention to the fragment deformation parameter $\delta$ under the
condition of small variation in $z$. Particulary in the synthesis
of the superheavy elements, when we treat the fusion process, the
fragility of the composite system with respect to $\delta$ is
crucial, because, as mentioned in Introduction, the saddle point
for fission is far inside the contact configuration, and the
instability due to the fragment deformation is easily realized.


\subsection{Characteristics of the mean trajectories}

First, we want to clarify how the critical condition changes upon
separating the trajectory for fusion and that for quasi-fission
when the atomic number of the compound nucleus increases. In order
to see the fundamental variation of the critical condition, we
employ the potential energy of the liquid drop model, and
calculate the mean trajectory. That is to say, we delete the final
term on the right-hand side of Eq.~(1). We want to exclude the
influence of the individuality of each nuclei. In this stage, the
shell effect is not taken into account considering that the shell
effect on the fusion probability is expected to cause the
fluctuation around the mean trajectory on the liquid drop
potential, and it is better to understand the fundamental
systematics of fusion hindrance. Actually, we discuss the fusion
probability including the shell effect in subsection 3.4.

Figure~2 shows the mean trajectory for forming the compound
systems $^{224}$Th, $^{232}$Pu, $^{240}$Cf, $^{256}$No,
$^{267}$Sg, $^{280}$Ds, $^{292}$114 and $^{297}$118, which are
produced by hot fusion reaction. For all systems, the entrance
channel is chosen to have the same mass asymmetry, $\alpha=0.6$,
and the incident energy corresponding to the excitation energy of
the compound nucleus $E^{*}=50$ MeV. The mean trajectory for $l=0$
is projected onto the $z-\alpha$ ($\delta=0$) and $z-\delta$
($\alpha=0$) planes in Figs.~2(a) and (b), respectively. The
trajectory calculation starts at the point of contact located at
$(z,\delta,\alpha)=(1.56, 0.0, 0.6)$ indicated by the arrow.


\begin{figure}
\centerline{
\includegraphics[height=.80\textheight]{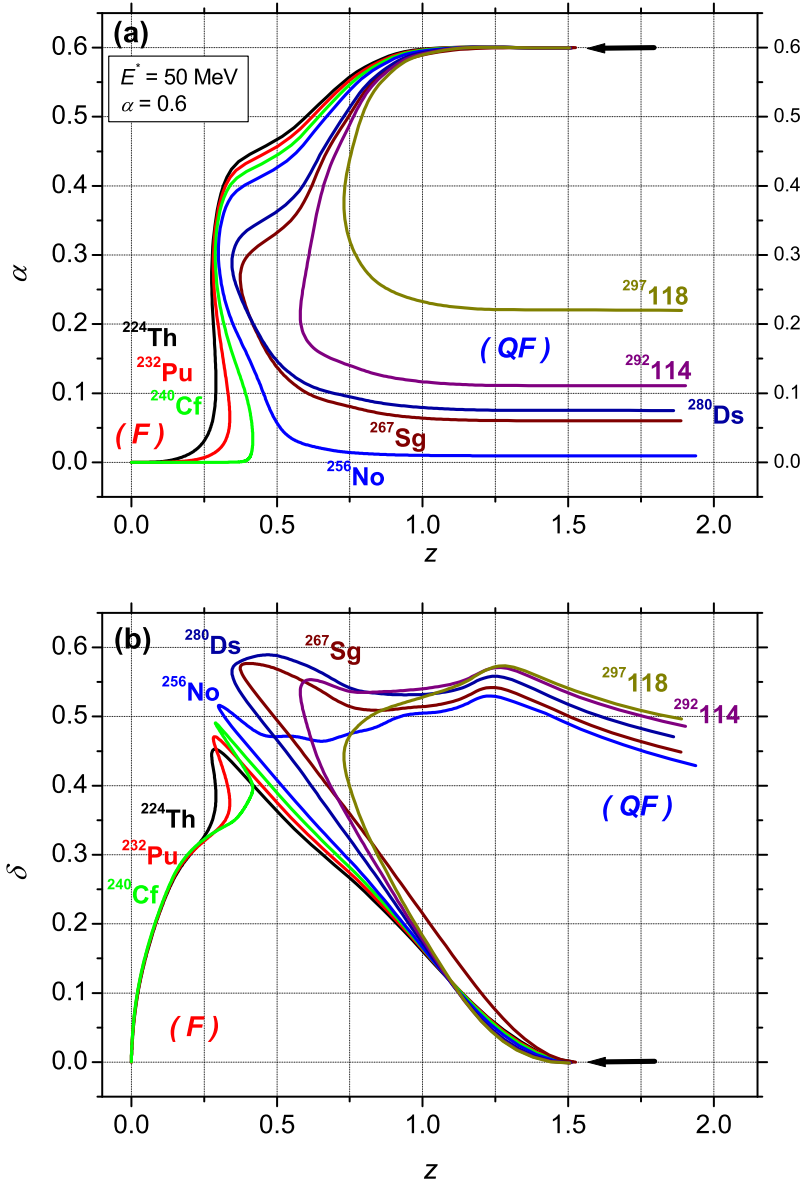}}
  \caption{The mean trajectories for forming various compound systems.
The entrance channels are chosen to have $\alpha=0.6$ and
$E^{*}=50$ MeV. The mean trajectories are projected onto the (a)
$z-\alpha$ ($\delta=0$) plane and (b) $z-\delta$ ($\alpha=0$)
plane. The starting point of the calculation is
$(z,\delta,\alpha)=(1.56, 0.0, 0.6)$, which is indicated by
arrows. The trajectory is classified into two groups: one leading
to fission (marked by F) and the other leading to quasi-fission
(marked by QF).}
\end{figure}



\begin{figure*}
\centerline{
\includegraphics[height=.32\textheight]{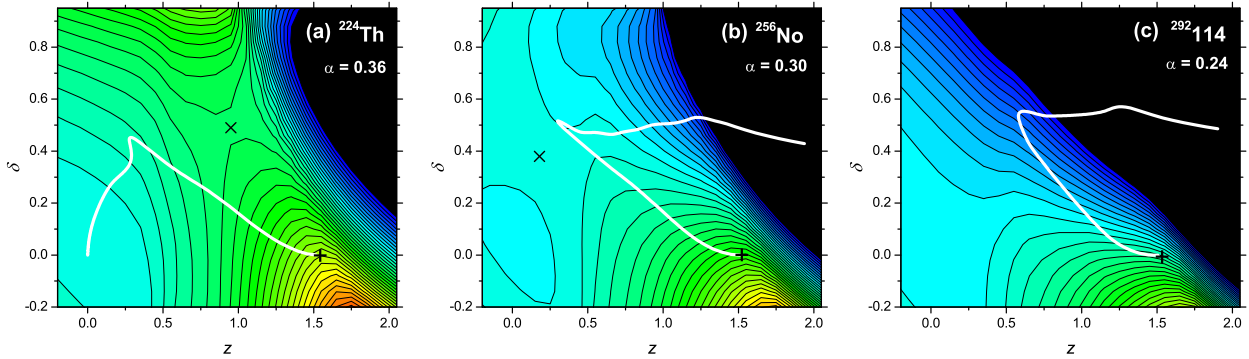}}
  \caption{The potential energy surfaces on the $z-\delta$ plane
for (a) $^{224}$Th, (b) $^{256}$No and (c) $^{292}$114. The mass
asymmetry parameter $\alpha$ is chosen as the value for each
critical area. The mean trajectory projected onto the $z-\delta$
plane is drawn by the white line. The initial point is marked by
+. We define the area where the direction of the trajectory
drastically changes as the critical area.}
\end{figure*}

The above-mentioned mean trajectories are grouped into two parts,
that is, the fusion (F) and the quasi-fission (QF) process. On the
$z-\alpha$ plane, for the trajectories of $Z=90, 94$ and 98
systems, the mass asymmetry parameter $\alpha$ relaxes quickly in
the region of small $z$, and ultimately, the trajectories go to
the spherical region. The trajectory of the $Z=102$ system at
first arrives in the area of $z\sim 0.3$, goes to the positive $z$
direction and then finally moves to mass symmetric fission. With
increasing atomic number of the system, it becomes increasingly
difficult for the trajectory to reach the smaller $z$ region and
it ultimately returns to the mass asymmetric fission region. We
can see the tendency that the mass asymmetry of fission fragments
increases with increasing atomic number of the reaction system,
which is consistent with the experimental data presented by Bock
et al. \cite{bock82}.

The situation can be understood by examining the characteristic
behavior of the trajectories projected onto the $z-\delta$ plane,
as shown in Fig.~2(b). As discussed in references
\cite{ari04,ari05}, the fragment deformation parameter $\delta$
plays a very important role in the fusion-fission process in the
superheavy mass region. Even for the $Z < 102$ system, the mean
trajectory at first goes to the positive $\delta$ direction but
the direction changes to the negative $\delta$ direction and
reaches to the spherical region.

All trajectories go to the positive $\delta$ direction from the
point of contact, and change their direction at around  $\delta
\sim 0.5 $. This trend is common for all systems treated here and
we call this area the {\it critical area}. In systematic
investigation, we can see that the behavior of the trajectory in
the critical area is the key in deciding whether it becomes the
fusion process or the quasi-fission process. The potential
landscape in the critical area is strongly related with the fusion
probability, as we discuss later.


\subsection{Critical condition of fusion process}

In order to understand the difference appearing in the
trajectories of fusion and quasi-fission processes, we again
investigate the landscape of the potential energy surface on the
$z-\delta$ plane. Figures~3(a), (b) and (c) show the potential
energy surface with the trajectory for forming $^{224}$Th,
$^{256}$No and $^{292}$114, respectively. These are drawn by the
cross section at the parameter $\alpha$ close to the critical area
which is indicated in the figure. The white line denotes the mean
trajectory and the initial point is marked by $(+)$.

For $^{224}$Th, the potential barrier develops well at large
$\delta$. In the critical area, the mean trajectory automatically
descends to the spherical region along the potential slope. We can
still see the saddle point marked by $(\times)$ in this system.
Here, the critical area for the mean trajectory is located just
inside the saddle point. Therefore, the fusion process is
predominant. On the other hand, for $Z > 102$ systems, the
critical area is outside the saddle point even if it is still
observed and no potential barrier can be seen at large $\delta$,
as shown in Figs.~3(b) and (c). The critical area is located
almost midway on the slope leading to fission.

The critical situation can clearly be seen when we plot the cross
section of the potential energy surface around the critical area.
Figure~4 shows the one-dimensional potential energy surface
depending on $\delta$, with fixed $z=0.3$ and $\alpha=0.3$.  The
gradient of the potential energy surface changes from positive to
negative at $Z=102$. This feature is the essential factor for the
mean trajectory whether it follows the fusion process or the
quasi-fission process.

As can be seen from Fig.3, the mean trajectory on the $z-\delta$
plane seems to stop for a moment at the critical area. The mass
asymmetry parameter $\alpha$ relaxes drastically at this moment,
as shown in Fig.~2(a). When the slope of the potential in the
critical area is relatively gentle, the trajectory spends a long
time here and the mass asymmetry easily relaxes. On the other
hand, when the slope in the critical area is steep toward the
fission region, the residence time of trajectory is too short for
relaxing the mass asymmetry completely, and the trajectory goes to
the mass asymmetric fission (or quasi-fission) region.


\begin{figure}
\centerline{
\includegraphics[height=.40\textheight]{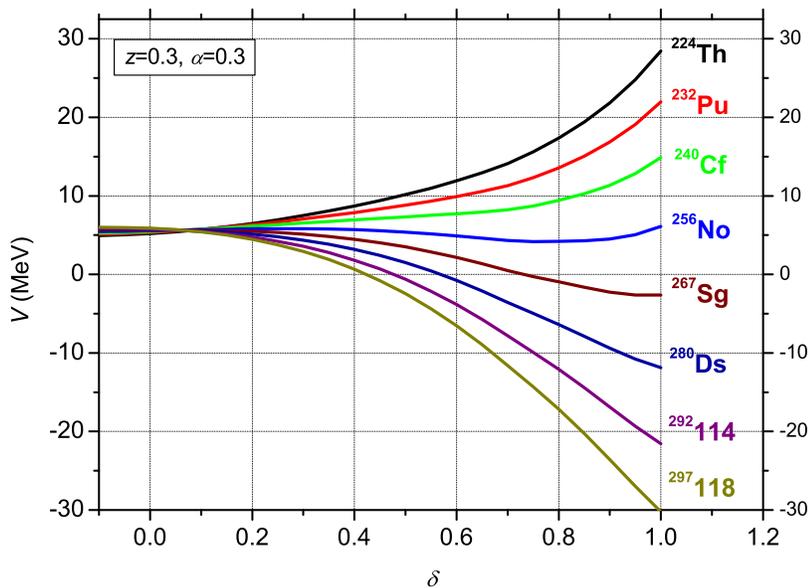}}
  \caption{One-dimensional potential energy surface of the various
systems depending on $\delta$, with fixed $z=0.3$ and
$\alpha=0.3$. It roughly corresponds to the cross section of the
potential energy surface in the critical area. The stability of
the composite nucleus against $\delta$ changes in $^{256}$No.}
\end{figure}



\subsection{Fusion probability forming superheavy elements}

Next, we discuss the systematics of the fusion probability using
the fluctuation-dissipation model. Using the same procedure as
described in reference \cite{ari04}, we calculate the fusion
probability taking into account the fluctuation around the mean
trajectory. Figure~5 shows the results for each system, in the
case of $l=0$. The calculations are performed for the case of two
entrance channel mass asymmetries, $\alpha=0.6$ and 0.0. In
Fig.~5(a), since the mean trajectories for $Z=90,94$ and 98
systems are classified as fusion processes, the fusion probability
is almost unity even if the fluctuation effect is taken into
account. For $Z
> 102$ systems, the fusion probability decreases exponentially with increasing
$Z$ number.


It is strongly related to the landscape of the potential energy
surface. With decreasing entrance channel mass asymmetry, the
fusion probability decreases due to the relationship between the
ridge line on the $z-\delta$ plane and the point where the kinetic
energy dissipates, as mentioned in reference \cite{ari05}. When
the initial value is $\alpha=0$, the fusion probability becomes
smaller, because the distance between the ridge line on the
$z-\delta$ plane and the point where the kinetic energy dissipates
is large.


\begin{figure}
\centerline{
\includegraphics[height=.55\textheight]{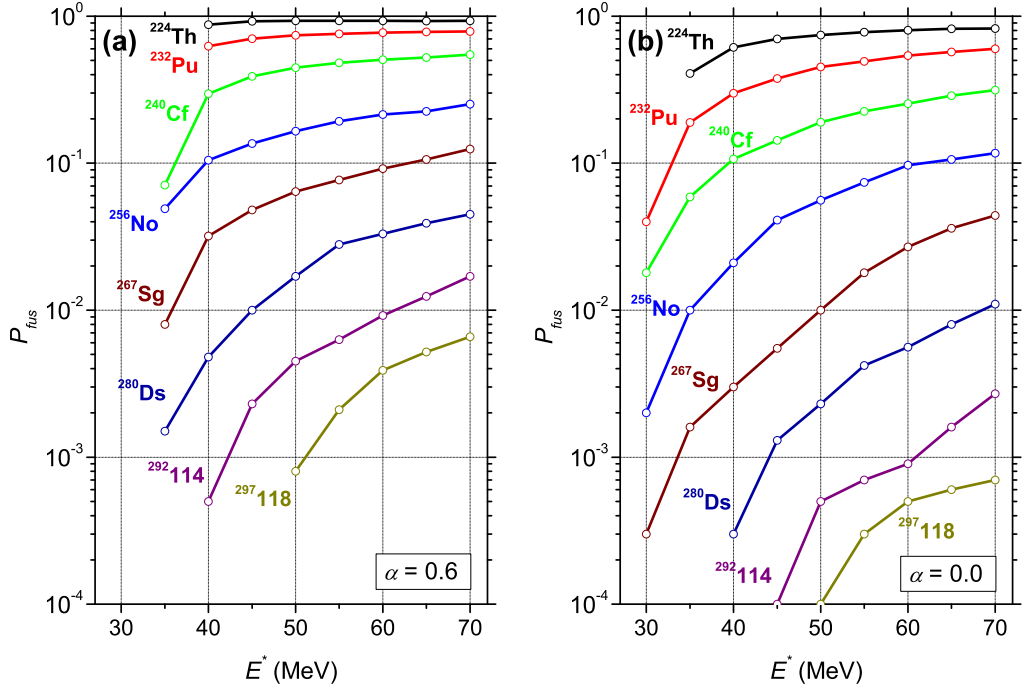}}
  \caption{Fusion probability for each system calculated using the
fluctuation-dissipation model, in the case of $l=0$. The initial
conditions are (a)~$\alpha=0.6$ and (b)~$\alpha=0.0$. }
\end{figure}


Though we discussed the general aspect of the fusion process using
the macroscopic model, in order to compare the calculations with
the experimental data, we should consider the shell effect in
fusion process, as we mentioned in Introduction. The shell effect
plays a very important role and enhances sometime the fusion
probability (fusion enhancement) \cite{gra99,moll97,armb99,gra00}.
Here, we calculate the fusion probability taking into account the
shell correction energy for the potential energy surface.

In the cold fusion reaction, Pb target is chosen to suppress the
excitation energy of compound nucleus \cite{oga75}. Due to the
strong shell structure of Pb target, the potential energy at the
contact point of the system is smaller than that of only the
liquid drop model. Moreover, the fusion valley ("cold valley")
which originates by the shell effect leads to enhance the fusion
probability \cite{gra99,gra00,ari04,ari05}. We calculate the
fusion probability for $l=0$ in the potential energy surface of
the $LDM$ and $LDM$ with the shell correction energy, which are
shown in Fig.~6 (a) and (b), respectively. As an initial
condition, we use the combination of the cold fusion reaction.
In the calculations shown in Fig.~6(b), we use the full shell
correction energy for the potential energy surface, though the
shell correction energy depends on the nuclear temperature $T$.
This is our intentional way to demonstrate the role of the shell
correction energy in the dynamical process.

When we take into account the shell correction energy in the
potential energy surface, the fusion probability increases for all
systems. The shell correction energy changes the potential
landscape, and it becomes easy for the trajectory to reach the
fusion region, which has been discussed precisely in reference
\cite{ari04,ari05}. For $Z < 102$ in Fig.~6, the fusion
probability shows the almost unit.



In order to calibrate our calculation, we compare our results with
the experimental data in the cold fusion reactions. Figure~7 shows
the fusion probability on Pb-target reaction series. We plot the
experimental data \cite{oga00} at the excitation energy
corresponding the Bass barrier height \cite{bass741} by open
squares. The calculation for $l=0$ is denoted by solid circles.
The fusion hindrance are seen from $Z_{1} \times Z_{2} > 1750 $,
where $Z_{1}$ and $Z_{2}$ denote the atomic number of the target
and projectile, and the fusion probability decreases drastically.
The tendency of the experimental data is reproduced by our
calculation.

We have discussed the mechanism of the dynamical process in the
case of $l=0$. When the system has an angular momentum, the
potential landscape changes, and the absolute value of the fusion
probability changes. However, the general tendency of the fusion
probability for each system is very similar in any angular
momentum cases.

\begin{figure}
\centerline{
\includegraphics[height=.55\textheight]{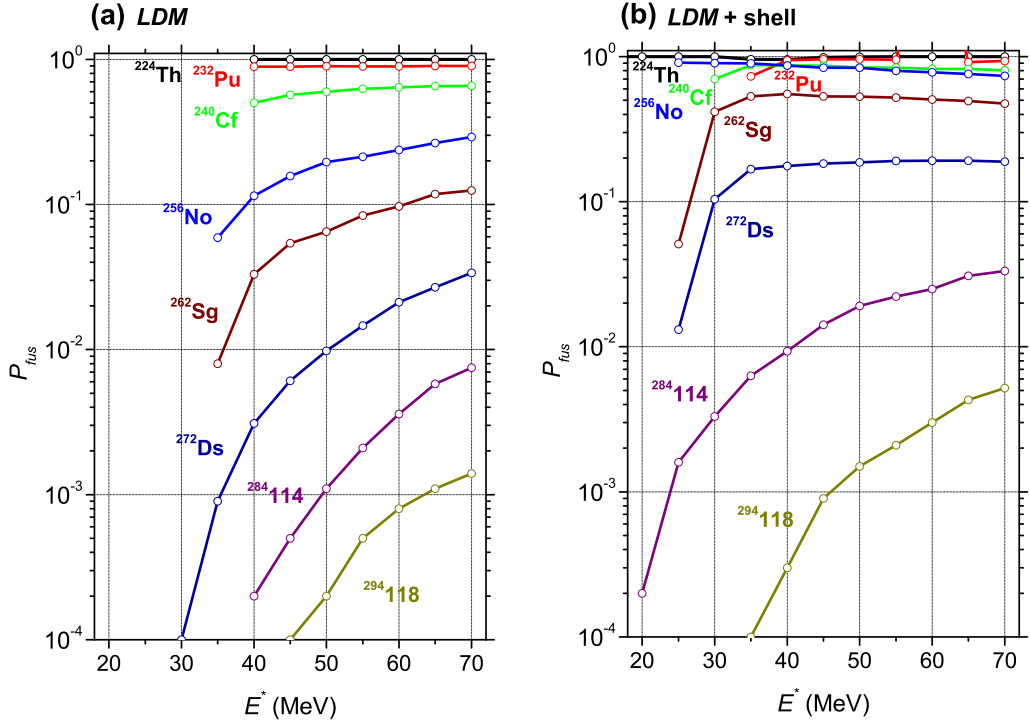}}
  \caption{Fusion probability in the cold fusion reactions calculated by the
fluctuation-dissipation model, in the case of angular momentum
$l=0$. The potential energy surface is (a) $LDM$ and (b)$LDM$ +
shell correction energy.}
\end{figure}

\begin{figure}
\centerline{
\includegraphics[height=.50\textheight]{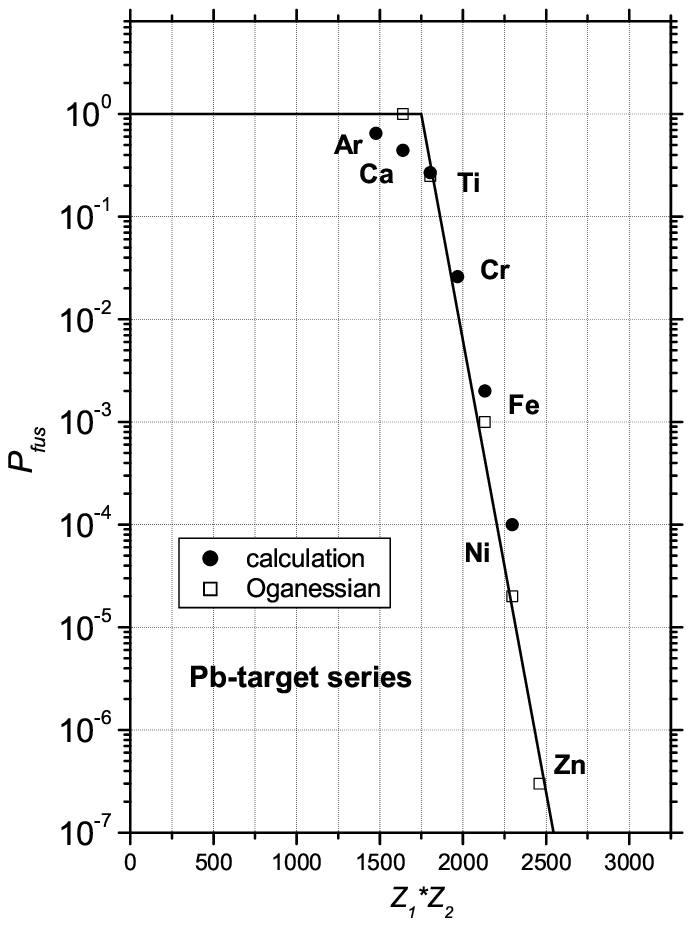}}
  \caption{Fusion probability for Pb-target reaction. We plot the experimental
  data at the excitation energy corresponding the Bass barrier height by open
  squares. The calculation in the case of angular momentum
$l=0$ is denoted by solid circles. The name of the element
  attached to the marks are the projectile.}
\end{figure}

\section{Summary}

The fusion-fission process in the superheavy mass region was
studied systematically by the trajectory calculation in the
three-dimensional coordinate space. In order to see the
systematics clearly, we employed the potential energy of the
liquid drop model, and calculated the mean trajectory. We
investigated the mechanism of the occurrence of fusion hindrance
by increasing the atomic number of the system. The mean trajectory
for all systems treated here at first goes to the positive
$\delta$ direction. Then, the gradient of the $\delta$ direction
of the potential energy surface in the critical area governs
whether the mean trajectory becomes the fusion or the
quasi-fission process. The fusion probability is consequently
estimated from the diffusion along the mean trajectory. From the
behavior of the mean trajectory shown in Fig.~3, we can understand
that the fusion probability decreases exponentially as the $Z$
number of the fused system increases. It is concluded that the
fusion hindrance in a system with $Z$ greater than 102 should be
described by considering the fragment deformation as an important
factor.

We assumed that each fragment has the same deformation to avoid
the consuming the four-dimensional calculation. We are recognizing
that the two deformation parameters are important, especially in
the fission process. We need further improvement on this point. It
is unclear the temperature dependence of the shell correction
energy for very deformed nuclei, especially near the point of
contact. We plan to study this problem, and to clarify the
influence of the temperature dependence of the shell correction
energy on fusion-fission process.


The authors are grateful to Professors Yu.~Ts.~Oganessian,
M.G.~Itkis, V.I.~Zagrebaev, F.~Hanappe and R.A.~Gherghescu for
their helpful suggestions and valuable discussion throughout the
present work. The authors thank Dr. S.~Yamaji and his
collaborators, who developed the calculation code for potential
energy with two-center parameterization. This work has been in
part supported by INTAS projects 03-01-6417.



\bibliographystyle{aipproc}   

\end{document}